\documentclass[aps,prd,floatfix,amsmath,amssymb,onecolumn,groupedaddress,12pt]{revtex4}
\usepackage{graphicx}
\usepackage{amssymb}
\usepackage{epstopdf}
\usepackage{subfigure}
\usepackage{array}
\usepackage{color}
\usepackage{pdfsync}
\usepackage{ragged2e}
\usepackage{amsmath}
\usepackage{hyperref}
\usepackage{epstopdf}
\begin{document}
\title{ Electroweak Naturalness and Deflected Mirage Mediation}
\author{Vernon Barger}
\email{barger@pheno.wisc.edu} 
\author{Lisa L.~Everett}
\email{leverett@wisc.edu} 
\author{Todd S.~Garon}
\email{tgaron@wisc.edu}  
\affiliation{University of Wisconsin - Madison}
\date{\today}

 \begin{abstract}
 We investigate the question of electroweak naturalness 
within the deflected mirage mediation (DMM) framework for supersymmetry breaking in the minimal supersymmetric standard model (MSSM).  The class of DMM models considered are nine-parameter theories that fall within the general classification of the 19-parameter phenomenological MSSM (pMSSM).   Our results show that these DMM models have regions of parameter space with very low electroweak fine-tuning, at levels comparable to the pMSSM.  These parameter regions should be probed extensively in the current LHC run.
 \end{abstract}
\maketitle
%

\section{Introduction}
Theories with TeV-scale supersymmetry, such as the minimal supersymmetric standard model (MSSM) and its extensions, have long been considered to be leading candidates for new physics that can elucidate the origin of electroweak symmetry breaking and address the gauge hierarchy problem associated with the Standard Model Higgs boson.   With the null result to date of searches for superpartners, and with the very recent turn-on of the Large Hadron Collider (LHC) which will probe TeV-scale energies at an unprecedented level,  the issue of theoretical ``naturalness" is a key question for this class of theories.   

Of course, the question of how ``fine-tuned" a specific model is depends on the criteria used to gauge it.  One general method, which has become a standard approach, is to evaluate the sensitivity to observables such as the $Z$ boson mass $m_Z$ to changes in the input parameters at a high scale, for example by the Barbieri-Giudice fine-tuning measure \cite{bgrefs},
\begin{eqnarray}
\Delta_{\rm BG}  = {\rm max}_i \left \vert \frac{\partial \ln m_Z^2}{\partial \ln a_i} \right \vert,
\end{eqnarray}
in which the $a_i$ represent the parameters at the theory.  This measure quantifies the extent to which electroweak to TeV scale observables are sensitive to variations in the high scale parameters, and as such is a gauge of the naturalness of the theory.  

However, it has recently been emphasized (see e.g.~\cite{Baer:2012uy}) 
that to address the specific question associated with naturalness in light of the non-observation of supersymmetry at the LHC, which is how does the observed value of $m_Z$ emerge when the superpartners must generically have masses that far exceed this value, fine-tuning measures other than $\Delta_{\rm BG}$ may yield valuable information.  One specific fine-tuning measure of this type is known as the electroweak fine-tuning measure $\Delta_{\rm EW}$, \cite{Baer:2012uy,Baer:2012mv,Baer:2012cf,Baer:2013gva,Baer:2013bba,Baer:2013xua,Baer:2013vpa,Baer:2013yha,Bae:2015nva,
Baer:2014ica,Baer:2015tva,Baer:2015rja,Bae:2015jea}, which assesses the extent to which cancellations occur in the prediction of the $Z$ boson mass as a function of the model parameters.  In practical terms, fine-tuning by this measure is a reflection of the degree to which the model parameters that enter in the expression for $m_Z$ are of order $m_Z$ themselves at the electroweak scale (low fine-tuning), or are much larger (high fine-tuning).  It has been emphasized previously that
the naturalness measure $\Delta^{-1}$ can serve as a Bayesian prior and as a likelihood estimate  \cite{Strumia:1999fr, Allanach:2006jc}.

More precisely, in the MSSM, the $Z$ boson mass is given at one loop by the following well-known relation:
\begin{eqnarray}
\frac{m_Z^2}{2}=\frac{m_{H_d}^2+\Sigma_d^d-\left(m_{H_u}^2+\Sigma_u^u\right)\tan^2\beta}{\tan^2\beta-1}-\mu^2,
\label{MZ}
\end{eqnarray}
in which the $\Sigma_{u,d}^{u,d}$ are the one-loop corrections for down-type and up-type quarks respectively (explicit expressions can be found in \cite{Baer:2012cf}).
The expression for $\Delta_{\rm EW}$ then takes the form
\begin{eqnarray}
\Delta_{\rm EW}={\rm max}_i \vert C_i \vert /(m_Z^2/2),
\label{deltaEWgen}
\end{eqnarray}
in which the $C_i$ are the terms in  Eq~\eqref{MZ}, for example $-m_{H_u}^2\tan^2\beta/(\tan^2\beta-1)$. As each of the $C_i$ are defined at the electroweak scale, each is determined purely by the supersymmetric spectrum, independent of the high-scale dynamics and renormalization group running effects that yield that spectrum.  For this reason, this fine-tuning assessment is often referred to as a determination of the degree of ``electroweak naturalness" of a given model.

 In studies of electroweak naturalness for various models of the MSSM soft terms, 
it has been noted that several general conditions at the electroweak scale result in low values of $\Delta_{\rm EW}$, and hence a small degree of fine-tuning.  These conditions include (i) $|\mu| \sim 100$ GeV, which results in light higgsino-like neutralinos, (ii) $m_{H_u}^2(m_Z)\sim -m_Z^2/2$ (as easily seen from Eq.~(\ref{MZ})), and (iii) large $A_t$, which is neeed to raise the Higgs mass without requiring heavy stops.  These conditions are not easily met within certain classes of models, but can be achieved in others.  In the 19-parameter phenomenological MSSM (pMSSM) \cite{Berger:2008cq} it is to be expected that there are regions of parameter space that meet these criteria and hence yield low values for $\Delta_{\rm EW}$, which was shown explicitly in a recent study \cite{Baer:2015rja}.  Similarly, this can also be achieved in the 19-parameter supergravity mode (SUGRA19) \cite{Baer:2013bba}.

However, in models with fewer free parameters, clearly these conditions are more difficult to achieve.    For example, minimal gauge mediation has difficulty because the $A$-terms are generated at two loops, and $m_{H_u}^2$ tends to run large and negative.   Models of mirage mediation (the mixed moduli-anomaly mediation scenario \cite{Choi:2005ge,Choi:2005uz,Endo:2005uy,Falkowski:2005ck} based on the Kachru-Kallosh-Linde-Trivedi (KKLT) construction  \cite{Kachru:2003aw}), also have difficulties, tending to have either large values of $\mu$, for similar reasons as minimal gauge mediation, or small values of $\mu$ and fail constraints on $B$ meson decays \cite{Baer:2014ica} (see \cite{Baer:2006tb,Baer:2007eh,Kaufman:2013oaa} for studies of the phenomenology of mirage mediation).   However, the variation on minimal supergravity known as NUHM2 \cite{Matalliotakis:1994ft}, a non-universal Higgs model that has six free parameters, can satisfy all of these criteria, yielding results for $\Delta_{\rm EW}$ as low as $\sim 5-10$  \cite{Baer:2013gva,Baer:2014ica,Baer:2015rja}.  The low fine-tuning allowed in this scenario is quite striking given that the NUHM2 is only a six-parameter model.  As such, it has been dubbed ``radiative natural supersymmetry" (RNS), wherein the MSSM and electroweak symmetry breaking arise naturally as the low energy limit of an underlying SUSY grand unified theory, and its phenomenological implications at the LHC and for dark matter physics have been thoroughly explored \cite{Baer:2012uy,Baer:2012cf,Baer:2013xua,Baer:2013yha,Baer:2013vpa,Baer:2015tva,Bae:2015jea,Bae:2015nva}.

The purpose of this paper is to explore the question of electroweak naturalness within a class of supersymmetric models known as deflected mirage mediation (DMM) models.  This framework is a natural extension of mirage mediation to include additional contributions from gauge mediation \cite{Everett:2008qy,Everett:2008ey}. In deflected mirage mediation, the gauge-mediated contributions to the soft supersymmetry breaking parameters can be comparable to the gravity-mediated and gauge-mediated contributions at the GUT scale, which in turn produces distinct phenomenology and a rich theory space for exploring current and future LHC supersymmetry searches, including examples of both simplified and compressed supersymmetric spectra \cite{Altunkaynak:2010tn,Abe:2014kla,Everett:2015dqa}. It is worth noting that the question of fine-tuning using high-scale fine-tuning measures such as $\Delta_{\rm BG}$ within the DMM framework has been explored \cite{Abe:2014kla}, particularly in light of the Higgs mass measurement at the LHC \cite{ATLAS:2013mma,Chatrchyan:2013lba,Aad:2015zhl}, though there is no prior fine-tuning study of DMM using the electroweak naturalness criterion.

As will be discussed in more detail shortly, the deflected mirage mediation framework, in its most general form, has a rich parameter space that can include regions that are outside the realm of the pMSSM (e.g., that predicts nonuniversal scalar masses for the first and second generations).  However, for phenomenological reasons, it is useful to consider only the subspace of DMM theory space that falls within the pMSSM guidelines.  Hence, we consider this restricted DMM parameter region within this paper.  This will allow for a straightforward comparison with the pMSSM.  We will demonstrate that within DMM models of this type, there are regions of parameter space with $\Delta_{\rm EW}$ as low as $\sim3.7$, {\it i.e.}, it is roughly equivalent to the best-case scenarios in SUGRA19 and slightly better than the best-case scenarios in the NUHM2/RNS scenario. (Here we note that direct comparisons with the pMSSM scan done in \cite{Baer:2015tva} are difficult as they likely have not sampled enough of the space to capture the low fine-tuned regions that SUGRA19, NUMH2, and DMM explore, which are all embeddable at low energy in the pMSSM.)

The paper is organized as follows. In the next section, we quickly review the soft terms in DMM and discuss the parameter space for the subspace of DMM models of interest in this paper. In Section III, we investigate the question of electroweak naturalness for this class of models, and show that there is a region of parameter space with extremely low values of the fine-tuning measure $\Delta_{\rm EW}$. We then summarize and conclude in Section IV.

\section{Overview of Deflected Mirage Mediation Models}

Deflected mirage mediation models are characterized by three classes of contributions to the soft supersymmetry breaking parameters.  As is the case in mirage models, there is a KKLT-like contribution to the soft masses that consists of tree-level supergravity contributions associated with a modulus field, as well as comparable anomaly mediation terms at a high scale, which is taken for simplicity to be the grand unification (GUT) scale $M_{\rm G}$.  Deflected mirage mediation scenarios also include gauge-mediated contributions, which take the form of a deflection of the soft terms at some messenger scale $M_{\rm mess}$. The messenger fields associated with the gauge mediation terms are typically taken to be $N$ vectorlike pairs of fundamental representations of $SU(5)$.   In these scenarios, the MSSM matter and Higgs fields are also each characterized by a modular weight label $n_i$ that appears in the respective K\"{a}hler potential terms for each of these fields. (To be more rigorous, the K\"{a}hler potential for the MSSM matter fields is taken to be diagonal in family space.  For the matter and Higgs fields, it takes the generally of the form $K\sim \sum_i \widetilde{K}_i \overline{\Phi}_i \Phi_i$, with $\widetilde{K}_i = (T+\overline{T})^{-n_i}$).

More explicitly, the high scale soft terms at $M_G$ take the form
\begin{eqnarray}
\label{gaugino1}
M_a(\mu = M_{\rm G})&=&M_0\left [1+\frac{g_0^2}{16\pi^2}b_a^\prime \alpha_m \ln \frac{M_P}{m_{3/2}}\right ] ,\\
\label{trilinear1}
A_i(\mu = M_{\rm G})&=&M_0 \left [(1-n_i)-\frac{\gamma_i}{16\pi^2} \alpha_m \ln \frac{M_P}{m_{3/2}}\right ] ,\\
\label{mssq1}
m_i^2(\mu = M_{\rm G})&=&M_0 ^2 \left [(1-n_i)-\frac{\theta'_i}{16 \pi^2} \alpha_m \ln \frac{M_P}{m_{3/2}} -\frac{\dot{\gamma}'_i}{(16\pi^2)^2}\left (\alpha_m \ln \frac{M_P}{m_{3/2}}\right )^2 \right ],
\end{eqnarray}
in which $m_{3/2}$ is the gravitino mass ($m_{3/2}$ generically exceeds $M_0$ by about a loop factor in magnitude, and thus is typically of order $10-100$ TeV).   Note that the physical trilinear terms are  $A_{ijk} y_{ijk}$, in which $A_{ijk} = A_i+ A_j + A_k$.

In the above expressions, $b_a^\prime = b_a+N$, in which $b_a$ are the one-loop beta functions for the gauge couplings ($b_{1,2,3}=(33/5,1,-3)$ in the MSSM).  The anomalous dimensions $\gamma^\prime_i =\gamma_i$ are given by $\gamma_i = 2 \sum_a g_a^2 c_a(\Phi_i) - (1/2)\sum_{lm} \vert y_{ilm} \vert^2$, in which the $y_{ijk}$ are the normalized MSSM Yukawa couplings, and the $c_a$ are the quadratic Casimirs.   The $\dot{\gamma}_i$'s are given by $\dot{\gamma}_i = 2\sum_a g_a^4 b_a c_a(\Phi_i)- \sum_{lm} \vert y_{ilm} \vert^2 b_{yilm}$, in which $b_{yilm}$ is the beta function of the Yukawa coupling $y_{ilm}$.  The quantities $\theta_i^\prime = \theta_i$ are given by $\theta_i= 4 \sum_a g_a^2 c_a(Q_i)-\sum_{ijk} \vert y_{ijk} \vert^2 (3-n_i-n_j-n_k).$  (Explicit expressions for these quantities can be found for example in Appendix A of \cite{Everett:2015dqa}.)  

The threshold contributions due to gauge mediation at $M_{\rm mess}$ take the form
\begin{eqnarray}
\label{threshgaug1}
\Delta M_a(\mu = M_{\rm mess})&=& -M_0 N\frac{g_a^2(M_{\rm mess})}{16\pi^2}   \alpha_m \left (1 +\alpha_g \right ) \ln \frac{M_P}{m_{3/2}} ,\\
\label{threshmssq1}
\Delta m_i^2(\mu = M_{\rm mess})&=&M_0^2\sum_a 2 c_a N
\frac{g_a^4 (M_{\rm mess}) }{(16\pi^2)^2} \left [\alpha_m  (1+\alpha_g)  \ln \frac{M_P}{m_{3/2}}\right ]^2.
\end{eqnarray}
Note that the trilinear terms do not receive threshold contributions at one-loop order, and hence these contributions are negligible.

From Eqs.~(\ref{gaugino1})-(\ref{threshmssq1}), we see that the model parameters for a general DMM scenario thus include:  (i) an overall mass scale $M_0$ associated with the tree-level supergravity mediation, (ii) a dimensionless parameter $\alpha_m$, which denotes the relative importance of anomaly mediation with respect to the tree-level gravity mediation (the KKLT scenario predicts $\alpha_m=1$), (iii) the number of messenger pairs $N$, (iv) the messenger scale $M_{\rm mess}$, (v) the dimensionless parameter $\alpha_g$, which denotes the relative importance of gauge mediation with respect to anomaly mediation, (vi) the modular weights $n_i$, (vii) the ratio of electroweak Higgs vacuum expectation values, and (viii) the sign of $\mu$. Here the standard procedure has been followed in which the model-dependent Higgs parameters $\mu$ and $b=B\mu$ are replaced by $\tan\beta$, $m_Z$, and the sign of $\mu$.

In a general DMM model of this type, the soft scalar mass-squared parameters have generation-dependent labels given by the possibility of family-dependent $n_i$ values, as well as the presence of Yukawa couplings in the $\theta_i'$ and $\dot{\gamma}_i^\prime$ quantities.  While the contributions to the anomalous dimensions, etc.~are typically negligible for all practical purposes for the first and second generations due to the hierarchical SM fermion masses, a general assignment of the modular weights $n_i$ can yield a sizable non-universal contribution.  For simplicity as well as phenomenological reasons,  we thus will always restrict ourselves to a subspace of DMM parameter space in which the matter fields all carry a universal modular weight $n_M$.  In addition, we will assume that the two electroweak Higgs fields also carry an independent modular weight $n_H$, which introduces an amount of non-universality.  

The DMM scenarios studied here thus have nine independent parameters (2 masses, 6 dimensionless parameters, and one sign): the mass scales $M_0$ and $M_{\rm mess}$, the dimensionless quantities $\alpha_m$, $\alpha_g$, the number of $SU(5)$ messenger pairs $N$, the modular weights $n_M$ and $n_H$,  $\tan\beta$, and ${\rm sign}\ \mu$.   We note that with this assumption regarding the modular weights, these scenarios represent a subset of the full 19-parameter pMSSM.

\section{Electroweak Naturalness in DMM Models}

In our analysis of electroweak naturalness in this class of DMM models, we use a subset of the dataset as studied in  \cite{Everett:2015dqa}. This dataset was determined as follows: for a randomly chosen mirage mediation point in the region $M_0\in[1,5]$ TeV, $\tan\beta\in[5,50]$, and $\alpha_m\in[0,2]$; we build a three-dimensional scan in the DMM parameter, scanning $\alpha_g \in [-1,1]$ in steps of 0.05, $\log_{10}\left[M_{\rm mess}/{\rm GeV}\right] \in[5,14]$ in unit steps, and $N \in [1,5]$ in unit steps. The modular weights $n_M$ and $n_H$ for the matter and Higgs fields, respectively, are allowed to vary independently  between 0 and 1 in half integer steps. The renormalization group (RG) equations were solved using a version of the package SOFTSUSY~3.3.9~\cite{Allanach:2001kg} that has been modified to account for the gauge mediation contributions~~\cite{Everett:2008qy,Everett:2008ey,Everett:2015dqa}. 

The phenomenological constraints applied to these model points are as follows. 
At the electroweak scale, points with negative mass-squares, or that do not result in electroweak symmetry breaking, or that do not have a neutralino LSP, are excluded. The surviving points are then cut according to an upper bound on the relic density,  $\Omega_{\chi}h^2\leq0.128$, taken from~\cite{Ade:2013zuv}, as calculated by MicrOMEGAs~2.2~\cite{Belanger:2001fz}, and a (conservative) Higgs mass bound  $123\,{\rm GeV}\leq m_h \leq 127\,{\rm GeV}$ \cite{ATLAS:2013mma,Chatrchyan:2013lba,Aad:2015zhl}. Finally, we apply constraints from $B_s\rightarrow \mu^+\mu^-$ and $b\rightarrow s \gamma$, with the values of  ${\rm Br}(B_s \to \mu^+ \mu^-)$ taken within $(1.5-4.3)\times10^{-11}$  \cite{Aaij:2012nna,Aaij:2013aka,Chatrchyan:2013bka} and a value of ${\rm Br}(b \to s\gamma)$ within $(3.03-4.08)\times10^{-4}$ \cite{Asner:2010qj}.  These cuts follow the ranges used in the previous work on $\Delta_{EW}$ \cite{Baer:2014ica,Baer:2015rja} to facilitate comparisons with these studies. In total, we use a~2 million  point subset and after application of all phenomenological constraints, leading to slightly more than 200,000 viable DMM model points.

\begin{figure}[t]
\begin{center}
\includegraphics[width=0.65\textwidth]{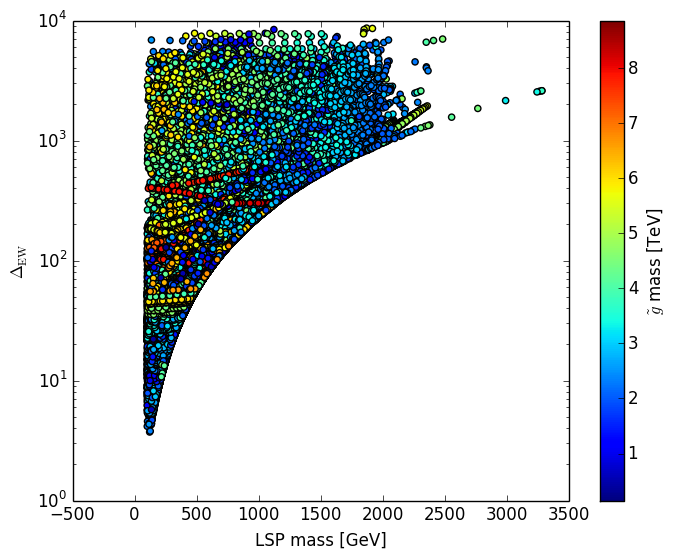}
\caption{$\Delta_{\rm EW}$ as a function of the LSP mass for the full dataset without a cut on the gluino mass, and shaded by the gluino mass in TeV.}
\label{plot:FullDist}
\end{center}
\end{figure}

\begin{figure}[t]
\begin{center}
\includegraphics[width=0.45\textwidth]{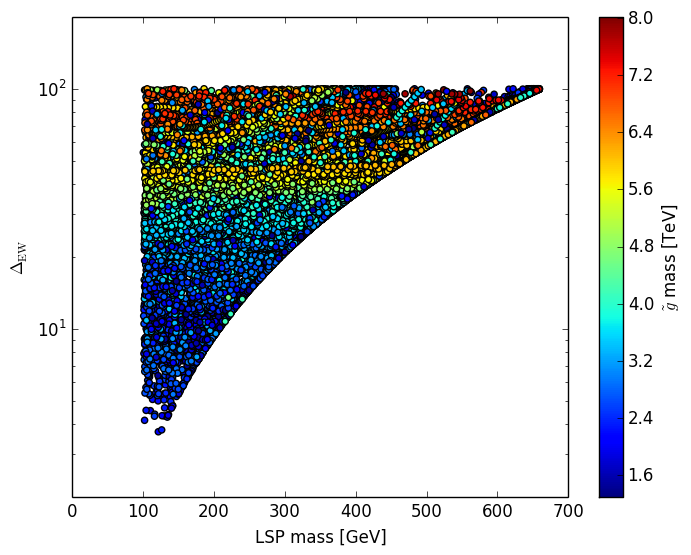}
\includegraphics[width=0.45\textwidth]{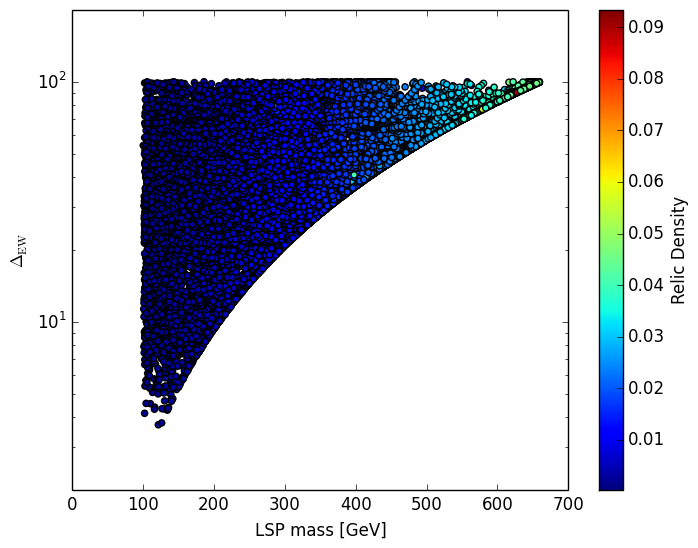}
\caption{A plot of $\Delta_{\rm EW}$ as a function of the LSP mass for points with $\Delta_{\rm EW}<100$, (left) shaded by the gluino mass in TeV, and (right) shaded by the relic density.}
\label{plot:LowDeltaDist}
\end{center}
\end{figure}
 
We now turn to the determination of $\Delta_{\rm EW}$ for these model points.  Figure~\ref{plot:FullDist} shows the results for $\Delta_{\rm EW}$ from the entire scan described previously and used in \cite{Everett:2015dqa}. The results of the figure do not change if we require $m_{\tilde g}>1.3$ TeV, consistent with current generic bounds from the LHC \cite{Aad:2015baa}, because the cut on ${\rm Br}(b\to s\gamma)$ removes some of the low mass gluinos. We use this bound on the gluino henceforth. 

In Figure~\ref{plot:FullDist} and the left panel of Figure~\ref{plot:LowDeltaDist}, we see that there is a large region that is less than 1\% fine-tuned. The minimum fine-tuning is of order 27\%, $\Delta_{\rm EM}\approx 3.7$. These are smaller than the minimum values for the NUHM2 model of $\Delta_{\rm EM}\approx 7$ \cite{Baer:2013gva,Baer:2015rja} or  $\approx10$ \cite{ Baer:2014ica}, slightly larger than the values found in SUGRA19 \cite{Baer:2013bba}. Furthermore, we see that if we want less than 1\% EWFT, then gluino masses less than about 8 TeV are allowed in DMM. 

\begin{figure}[t]
\begin{center}
\includegraphics[width=0.45\textwidth]{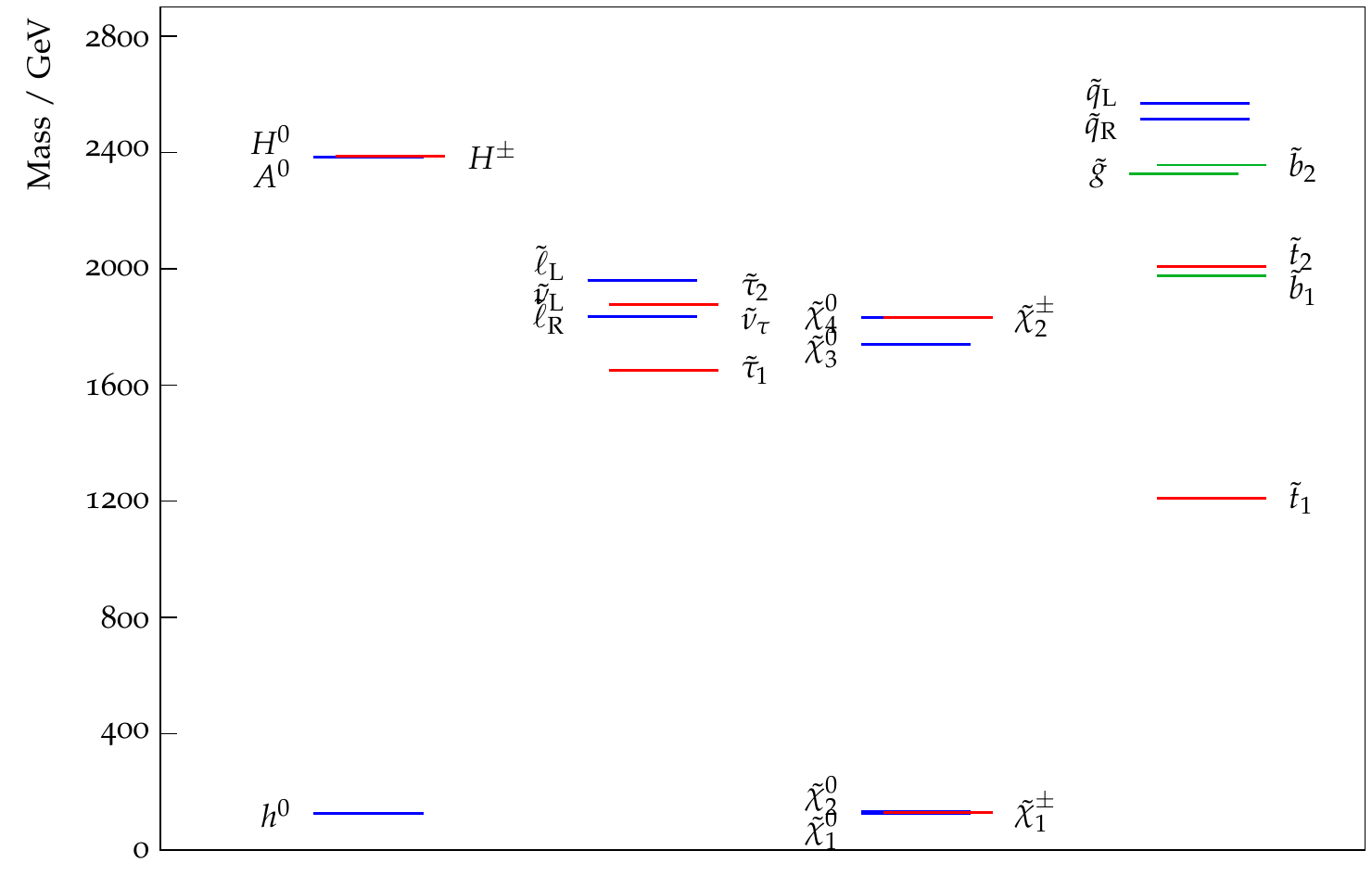}
\includegraphics[width=0.45\textwidth]{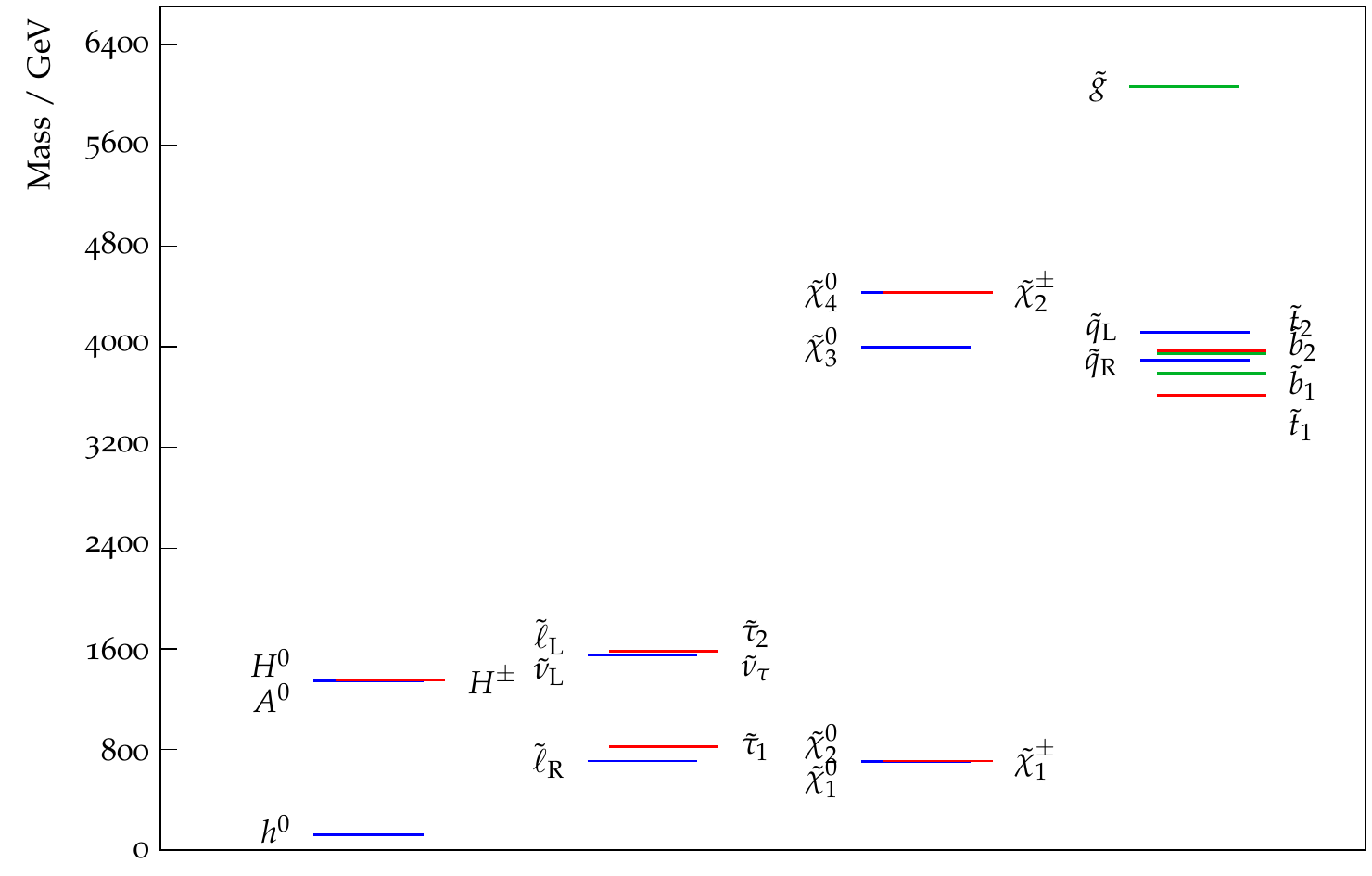}

\caption{Examples of the Higgs and superpartner spectra for two representative points with small(ish)  values of $\Delta_{\rm EW}$.  The left panel is a point with $\Delta_{\rm EW}<4$ and a low value of the relic density ($\Omega_\chi h^2= 2.4 \times 10^{-3}$), for which $M_0=2600$ GeV, $\alpha_m=1.21$, $\tan\beta=22$, $(n_M,n_H)=(0.5,0)$, $\alpha_g=0.1$, $M_{\rm Mess}=10^9$ GeV, and $N=2$.  The right panel is a point with $\Delta_{\rm EW}<120$ and a large value of the relic density ($\Omega_\chi h^2=0.113$), for which $M_0=4800$ GeV, $\alpha_m=1.36$, $\tan\beta=38$, $(n_M,n_H)=(1,1)$, $\alpha_g=-0.4$, $M_{\rm Mess}=10^{14}$ GeV, and $N=1$.}
\label{plot:LowDeltaSpectra}
\end{center}
\end{figure}

The right panel of Figure \ref{plot:LowDeltaDist}, shows the relic density as a function of the LSP mass. The points with the smallest fine-tuning typically have $O({150\, {\rm GeV}})$ LSPs and  do not fulfill the relic density constraint, and another non-thermal species such as an axion is needed \cite{Bae:2015jea}. This does not need to be the case if there are co-annihilations between a heavier  higgsino-like LSP and the right-handed sleptons. An example of this sort of spectra is shown in the right panel of Figure \ref{plot:LowDeltaSpectra}, in which less than 0.5\% fine-tuning can be achieved in a corner of parameter space with large $M_0$ and $\alpha_m$, and both modular weights equal to one. Removing the upper bound on the amount of fine-tuning admits spectra where the proper relic density can be achieved through co-annihilation with a stop or gluino as well. Over much of the higgsino-like parameter space, the difference between $\tilde \chi_2^0-\tilde \chi_1^0$ is typically $<10$ GeV and often less than $<5$ GeV, leading to very soft and likely hard to detect signals at LHC13 \cite{Baer:2013xua}.

The region with $\Delta_{\rm EW}$ less than $\sim 100$ is made up of points with mass spectra similar to the two example spectra shown in Figure~\ref{plot:LowDeltaSpectra}. These two points share a light, highly mixed stop and very light pure higgsino LSP, but they also share a near degeneracy among the next heaviest particles after the higgsino-like neutralinos and charginos. In the left panel, we have a near degenerate gluino and stop, and in the right a near degenerate stau, smuon, and selectron. As mentioned earlier, co-annihillation with the stau allows the spectrum to generate a value of the dark matter relic density near the measured value from the Planck experiment \cite{Ade:2013zuv}. The left-hand spectra in Figure~\ref{plot:LowDeltaSpectra} is similar to the spectra found in points with low fine-tuning in the NUHM2 model. In this model, and similarly in the SUGRA19 model, the least tuned points tend to have a $\sim 1$ TeV, highly mixed stop and a plethora of particles above $\sim 1.5$ TeV, including a $\sim 2$ TeV gluino. Unlike NUMH2, DMM does not have universal gaugino masses and so $M_1$ and $M_2$ can and tend to be much larger.

If we further use a tighter bound of $3.3\%$ fine-tuning, we can compare the bounds in the NUHM2 model from \cite{Baer:2015rja}. In most respects the two models agree. In DMM, the gluino mass is capped at about $5$ TeV, larger than the upper bound in \cite{Baer:2015rja}. Similarly the bounds on $M_1$ and $M_2$ of 900 and 1700 GeV respectively from  \cite{Baer:2015rja}, are instead 4.35 and 4.15 TeV in DMM. Non-universal gaugino masses, like those in DMM Eq \eqref{gaugino1}, relax the limits in mSUGRA-like models. Other bounds, like those on the lightest stop, are similar with NUHM2 and below the upper bounds from the pMSSM or SUGRA19 from \cite{Baer:2013bba,Baer:2015rja}. The results indicate that as we increase the number of degrees of freedom the bounds on particle masses, other than the lightest stop, and parameters, other than $\mu$ and $m_{H_u}^2$, weaken, but are still within the range of future colliders and Higgs factory-like experiments \cite{Cohen:2013xda}. 

Wino- and bino-like LSPs are typically more fine-tuned than higgsino-like points. The minimum point with a wino-like LSP has $\Delta_{\rm EW}\approx60$ and the minimum bino-like point in the sample has $\Delta_{\rm EW}\approx176$. These are both more fine-tuned than the values explored in the natural supersymmetry phenomenological study in the NUHM2 model of \cite{Baer:2015tva}, because they allow for other hierarchies in the soft terms other than the $M_1>M_2>M_3$ in DMM at the GUT scale. The messenger scale contribution Eq \eqref{threshmssq1} preserves the same hierarchy in gaugino masses, although the hierarchy in terms of the absolute value of the gaugino masses may change. This allows for wino-like LSPs while there were none in mirage mediation. For wino-like points, the deflection must be large, $N\geq 3$ and $\alpha_g\sim 1$, and at a low scale. The deflection for the Higgs masses are large, positive and leads to large $m^2_{H_u}$ at $M_{\rm SUSY}$, requiring a larger value of $\mu$ or the corrections to compensate, either solution leads to larger fine-tuning. If we were to allow a lower messenger scale or larger $\alpha_g$, we would likely admit wino-like DMM points with smaller fine-tuning. Running may also modify this hierarchy, opening up the possibility of regions with bino-like LSPs. Light, pure-bino dark matter tends to be over produced, setting a lower bound on $M_1$ and $\mu$, leading to increased fine-tuning in the bino-like sample.

\begin{figure}[t]
\begin{center}

\includegraphics[width=0.45\textwidth]{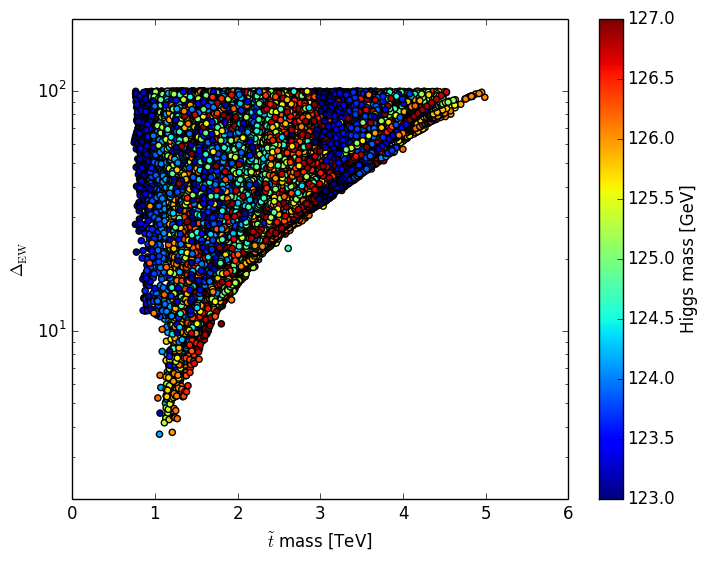}
\includegraphics[width=0.45\textwidth]{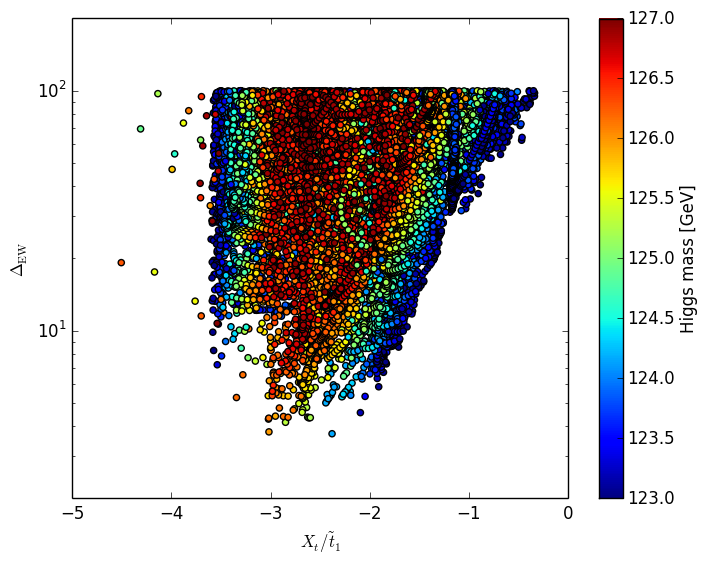}
\caption{A plot of $\Delta_{\rm EW}$ as a function of the stop mass (left) and $\Delta_{\rm EW}$ vs. $X_t$ divided by the stop mass (right), with $\Delta_{\rm EW}<100$ and a 1.3 TeV cut on the gluino mass. The notches in the lower left corners correspond to regions that are excluded by ${\rm Br}(b\rightarrow s \gamma)$ and ${\rm Br}(B_s\rightarrow \mu^+\mu^-)$.}
\label{plot:stopmix}
\end{center}
\end{figure}

An investigation of the Higgs mass as a function of the stop mass and the stop mixing parameter $X_t$, as shown in Fig \ref{plot:stopmix}, demonstrates that points with low fine-tuning are typically those that are near maximally mixed $|X_t/\tilde t_1|\sim 2.5$ and have TeV scale stops \cite{Baer:2011ab,Brummer:2012ns}. In the left panel, we see that in the $X_t/m_{\tilde t_1}$ versus $\Delta_{\rm EW}$ plane.  a lighter Higgs will typically allow points with lower fine-tuning. The notch in the lower left corner of both distributions corresponds to a region that is excluded by the constraint on ${\rm Br}(b\to s \gamma)$.

\begin{figure}[t]
\begin{center}

\includegraphics[width=0.6\textwidth]{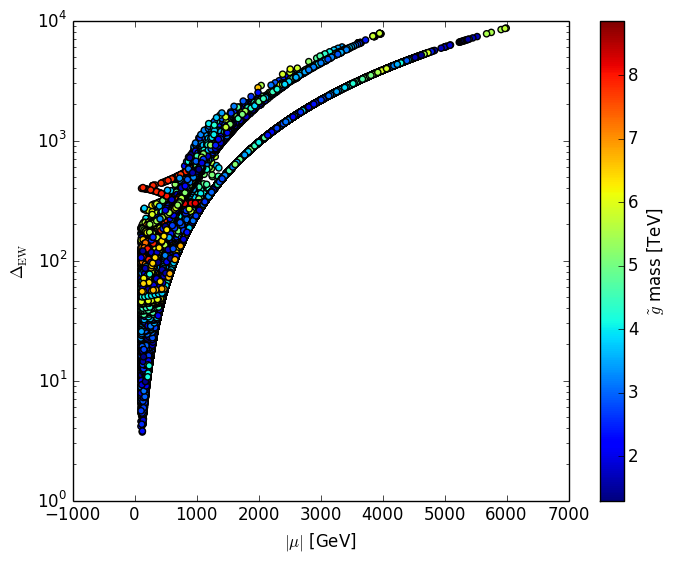}
\caption{A plot of $\Delta_{\rm EW}$ vs. the $\mu$ parameter, colored by the gluino mass. The two tails correspond to points in which $m_{H_u}^2$ is positive  (top) and negative (bottom) at $M_{\rm SUSY}$.}
\label{plot:mu}
\end{center}
\end{figure}

In Figure \ref{plot:mu}, we see that there are two distinctive regimes for the $\mu$ parameter.  In  one regime, $m_{H_u}^2$ is negative at $M_{\rm SUSY}$, and in the other,  $m_{H_u}^2$ is positive and runs to negative values below $M_{\rm SUSY}$. For Higgsino-like points, the latter set forms a tight band, while the former has a spread, but for the same value of $\mu$ points where $m_{H_u}^2$ is already negative are less fine-tuned than points where it is positive, but the positive points reach lower overall values of $\mu$ and of $\Delta_{\rm EW}$. The gap between the two branches comes from the dearth of points where $m_{H_u}^2\sim 0$ at values of $\mu\gtrsim 1$ {\rm TeV}, above which electroweak symmetry breaking is difficult to achieve. For bino-like points, all have negative $m_{H_u}^2$, with lower fine-tuning occuring for points with larger values of the up-type Higgs mass. Wino-like points arise from all values of $m_{H_u}^2$, with positive values typically having a smaller $\mu$ parameter and lower fine-tuning.

\begin{figure}[t]
\begin{center}
\includegraphics[width=0.45\textwidth]{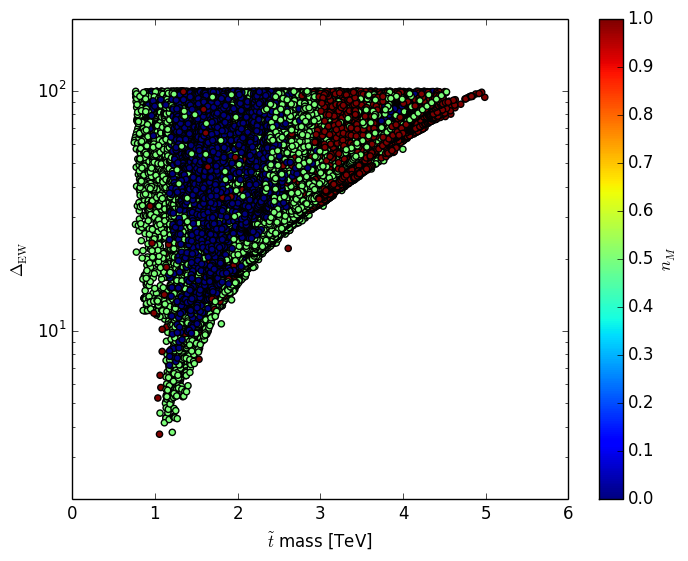}
\includegraphics[width=0.45\textwidth]{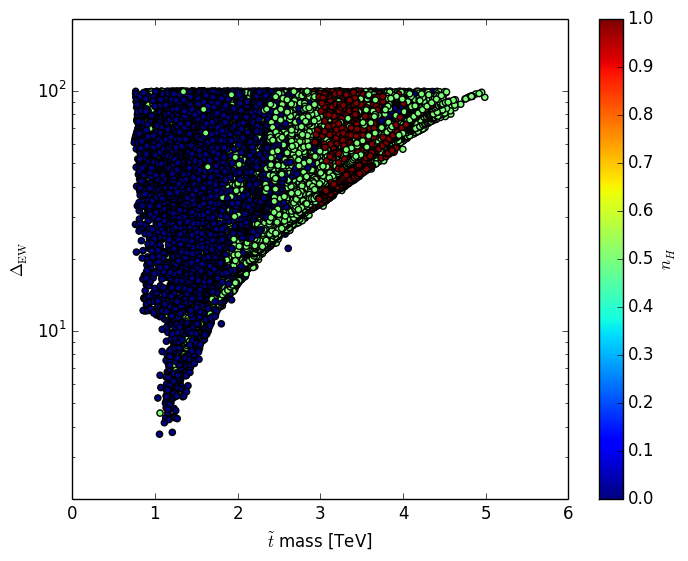}
\caption{$\Delta_{\rm EW}$ vs. the stop mass, with the values of the matter modular weight $n_M$ indicated on the left, and of the Higgs modular weight $n_H$ on the right, for points with $\Delta_{\rm EW}<100$.}
\label{plot:stopweights}
\end{center}
\end{figure}

If we break the sample down by modular weights as shown in Figure~\ref{plot:stopweights}, we see that the majority of points with extremely low fine-tuning have $n_H=0$. $n_H=0$ implies that the tree-level supergravity contribution to the Higgs soft mass-squared parameters is maximized (see Eq~(\ref{mssq1})), leading to small masses at the GUT scale, since the anomaly contribution has the opposite sign. In mirage mediation models, $m^2_{H_u}$ will typically run to large and negative, leading to large values of $\Delta_{\rm EW}$. In DMM, the addition of the messenger fields deflect the soft masses upwards, leading typically to shallower values at $M_{\rm SUSY}$ compared to mirage mediation. This is the radiatively natural scenario explored in \cite{Baer:2012cf,Baer:2013xua,Baer:2013yha,Baer:2013vpa,Baer:2015tva,Bae:2015jea,Bae:2015nva}. Since the corrections can be large, there are regions in DMM, where for $\Delta_{\rm EW}\gtrsim 7$, $n_H=1/2$ can lead to low fine-tuning as well. 

Our results show that in DMM, all values of $n_M$ studied can result in low electroweak fine-tuning, as opposed to the case of mirage models, which single out $n_M=1$ \cite{Baer:2014ica}. That being said, the best results in DMM tend to occur for $n_M=1$ or $1/2$.  These DMM points tend to have soft mass-squared parameters that are negative at the GUT scale but are positive at $M_{\rm SUSY}$ through RG evolution and the positive messenger scale deflection due to the gauge mediation terms. The addition of messengers in DMM leads to larger values for the gauge couplings at the GUT scale, causing the the anomaly mediation contribution to become increasingly large and negative, and so it may require a nonzero tree-level gravity contribution, $n_M\neq 1$, to moderate the soft mass- squared parameters to get light, $\mathcal O(1 {\rm TeV})$ stops.   In mirage models, the GUT scale value of the couplings is smaller, leading to a small positive anomaly mediation contribution at the GUT scale, which then runs to give us light stops. If $n_M\neq 1$ in these models, the addition of moduli would lead to a heavy stop and the model would be fine-tuned. 

Breaking the results down by other parameters reveal a few other trends in the fine-tuning results. DMM points with low fine-tuning tend to have one or two messengers which integrate out at a possibly high scale with potentially any value of $\alpha_g$.  Large values of $N$ also tend to deflect $m^2_{H_u}$ too much to run to a shallow minimum over much of the parameter space, giving large fine-tuning. Furthermore, $M_0$ must be greater than roughly $1.5$ TeV, to get mixed stops to bolster the Higgs mass, which leads to gluino masses between 2 and 4 TeV that are accessible at the LHC.

\section{conclusion}

\begin{figure}[t]
\begin{center}
\includegraphics[width=\textwidth]{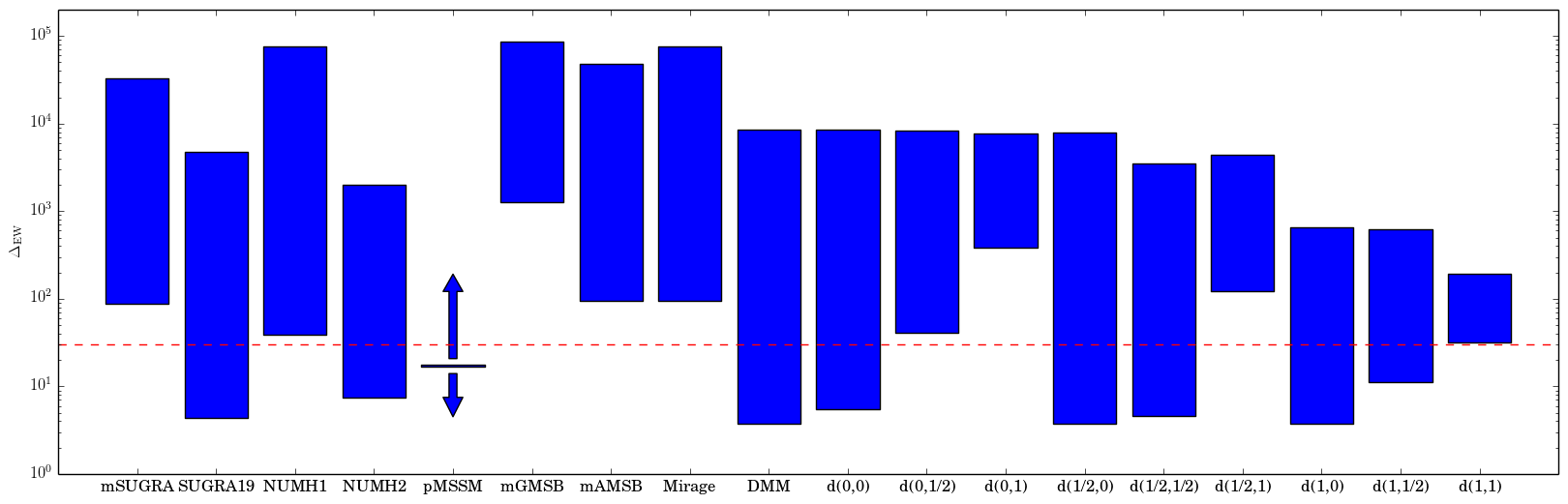}
\caption{A comparison of our results with the fine-tuning ranges found in many SUSY models as taken from \cite{Baer:2013bba,Baer:2014ica,Baer:2015rja}.  The line for the pMSSM denotes the lower bound determined in \cite{Baer:2015rja}; the arrows denote that the fine-tuning ranges may go below as well as above this line if a more comprehensive scan is performed.  The DMM points are further broken down by the modular weights $n_M$ and $n_H$ for the matter and Higgs fields; these points are denoted by $d(n_M, n_H)$. The dashed line represents $\Delta_{\rm EW}<30$, which is considered not fine-tuned.}
\label{plot:barchart}
\end{center}
\end{figure}

In this paper, we have shown that that deflected mirage mediation, a nine-parameter scenario in which gravity, anomaly, and gauge mediation all can contribute comparably to the soft supersymmetry breaking parameters of the MSSM,  admits spectra with low electroweak fine-tuning. 
A comparison of the electroweak fine-tuning ranges in DMM with many standard models for the soft terms of the MSSM is shown in Figure \ref{plot:barchart}.  Here we note that for the pMSSM, as an upper bound was not determined in \cite{Baer:2015rja}.  Hence, here we quote the lowest $\Delta_{\rm EW}$ presented (the line) and use the arrows to denote that the range of fine-tuning likely goes far up and down, past models that can be embedded in the pMSSM like mGMSB, and at least as low as models like DMM. Given the large parameter space of the pMSSM, we suspect that a more thorough scan of this scenario would lead to points that are at least, or less, fine-tuned than DMM or SUGRA19.

The results show that DMM does better in general than these standard scenarios, and is comparable to high-scale models with many more parameters such as SUGRA19. 
We saw that in DMM models with low numbers of messengers can lead to values with low fine-tuning. The other parameters that enter into the deflection, $\alpha_g$ and $M_{\rm mess}$, do not have any preferred values. For the parameters that enter into the GUT scale masses, we notice $\alpha_m>1$, with some points near $\alpha_m=2$, $M_0>1500$ GeV, and with any value of $\tan\beta$. Similarly $n_H=1$ almost exclusively leads to high fine-tuning, but other combinations of modular weights lead to acceptable values of $\Delta_{\rm EW}$. 
Hence, we see that in DMM models, the combination of gravity mediation, anomaly mediation, and gauge mediation opens up new  avenues with lower fine-tuning and should motivate us to look at models beyond the minimal set, where correlations between parameters can lead to unexpected results. 

\begin{acknowledgments}

We thank H.~Baer for useful discussions.  This work is supported by the U. S. Department of Energy under the contract DE-FG-02-95ER40896. 
\end{acknowledgments}


\begin{thebibliography}{99}



\bibitem{bgrefs}
  J.~R.~Ellis, K.~Enqvist, D.~V.~Nanopoulos and F.~Zwirner,
  Mod.\ Phys.\ Lett.\ A {\bf 1}, 57 (1986).
  doi:10.1142/S0217732386000105
  R.~Barbieri and G.~F.~Giudice,
  Nucl.\ Phys.\ B {\bf 306}, 63 (1988).
  G.~W.~Anderson and D.~J.~Castano,
  Phys.\ Lett.\ B {\bf 347}, 300 (1995)
  doi:10.1016/0370-2693(95)00051-L
  [hep-ph/9409419].
\bibitem{Baer:2012uy} 
  H.~Baer, V.~Barger, P.~Huang and X.~Tata,
  JHEP {\bf 1205}, 109 (2012)
  [arXiv:1203.5539 [hep-ph]].

\bibitem{Baer:2012mv} 
  H.~Baer, V.~Barger, P.~Huang, D.~Mickelson, A.~Mustafayev and X.~Tata,
  Phys.\ Rev.\ D {\bf 87}, no. 3, 035017 (2013)
  [arXiv:1210.3019 [hep-ph]].

\bibitem{Baer:2012cf} 
  H.~Baer, V.~Barger, P.~Huang, D.~Mickelson, A.~Mustafayev and X.~Tata,
  Phys.\ Rev.\ D {\bf 87}, no. 11, 115028 (2013)
  [arXiv:1212.2655 [hep-ph]].

\bibitem{Baer:2013yha} 
  H.~Baer, V.~Barger, P.~Huang, D.~Mickelson, A.~Mustafayev, W.~Sreethawong and X.~Tata,
  Phys.\ Rev.\ Lett.\  {\bf 110}, no. 15, 151801 (2013)
  [arXiv:1302.5816 [hep-ph]].
  
\bibitem{Baer:2013vpa} 
  H.~Baer, V.~Barger and D.~Mickelson,
  Phys.\ Lett.\ B {\bf 726}, 330 (2013)
  [arXiv:1303.3816 [hep-ph]].

\bibitem{Baer:2013bba} 
  H.~Baer, V.~Barger and M.~Padeffke-Kirkland,
  Phys.\ Rev.\ D {\bf 88}, 055026 (2013)
  [arXiv:1304.6732 [hep-ph]].
  
\bibitem{Baer:2013gva} 
  H.~Baer, V.~Barger and D.~Mickelson,
  Phys.\ Rev.\ D {\bf 88}, no. 9, 095013 (2013)
  [arXiv:1309.2984 [hep-ph]].

\bibitem{Baer:2013xua} 
  H.~Baer, V.~Barger, P.~Huang, D.~Mickelson, A.~Mustafayev, W.~Sreethawong and X.~Tata,
  JHEP {\bf 1312}, 013 (2013)
  [JHEP {\bf 1506}, 053 (2015)]
  [arXiv:1310.4858 [hep-ph]].
  
\bibitem{Baer:2014ica} 
  H.~Baer, V.~Barger, D.~Mickelson and M.~Padeffke-Kirkland,
  Phys.\ Rev.\ D {\bf 89}, no. 11, 115019 (2014)
  [arXiv:1404.2277 [hep-ph]].
 

 
\bibitem{Baer:2015tva} 
  H.~Baer, V.~Barger, P.~Huang, D.~Mickelson, M.~Padeffke-Kirkland and X.~Tata,
  Phys.\ Rev.\ D {\bf 91}, no. 7, 075005 (2015)
  [arXiv:1501.06357 [hep-ph]].
  
\bibitem{Bae:2015jea} 
  K.~J.~Bae, H.~Baer, V.~Barger, M.~R.~Savoy and H.~Serce,
  Symmetry {\bf 7}, no. 2, 788 (2015)
  [arXiv:1503.04137 [hep-ph]].
\bibitem{Bae:2015nva} 
  K.~J.~Bae, H.~Baer, N.~Nagata and H.~Serce,
  Phys.\ Rev.\ D {\bf 92}, no. 3, 035006 (2015)
  [arXiv:1505.03541 [hep-ph]].
  
\bibitem{Baer:2015rja} 
  H.~Baer, V.~Barger and M.~Savoy,
  arXiv:1509.02929 [hep-ph].
  
\bibitem{Strumia:1999fr} 
  A.~Strumia,
  [hep-ph/9904247].
\bibitem{Allanach:2006jc} 
  B.~C.~Allanach,
  Phys.\ Lett.\ B {\bf 635}, 123 (2006)
  doi:10.1016/j.physletb.2006.02.052
  [hep-ph/0601089].
  M.~E.~Cabrera, J.~A.~Casas and R.~Ruiz de Austri,
  JHEP {\bf 0903}, 075 (2009)
  doi:10.1088/1126-6708/2009/03/075
  [arXiv:0812.0536 [hep-ph]].
  S.~Cassel, D.~M.~Ghilencea, S.~Kraml, A.~Lessa and G.~G.~Ross,
  JHEP {\bf 1105}, 120 (2011)
  doi:10.1007/JHEP05(2011)120
  [arXiv:1101.4664 [hep-ph]].
  D.~M.~Ghilencea and G.~G.~Ross,
  Nucl.\ Phys.\ B {\bf 868}, 65 (2013)
  doi:10.1016/j.nuclphysb.2012.11.007
  [arXiv:1208.0837 [hep-ph]].
  
 
  
\bibitem{Berger:2008cq} 
  C.~F.~Berger, J.~S.~Gainer, J.~L.~Hewett and T.~G.~Rizzo,
  JHEP {\bf 0902}, 023 (2009)
  [arXiv:0812.0980 [hep-ph]].
 


\bibitem{Choi:2005ge} 
  K.~Choi, A.~Falkowski, H.~P.~Nilles and M.~Olechowski,
  Nucl.\ Phys.\ B {\bf 718}, 113 (2005)
  [hep-th/0503216].

\bibitem{Choi:2005uz} 
  K.~Choi, K.~S.~Jeong and K.~i.~Okumura,
  JHEP {\bf 0509}, 039 (2005)
  [hep-ph/0504037].

\bibitem{Endo:2005uy} 
  M.~Endo, M.~Yamaguchi and K.~Yoshioka,
  Phys.\ Rev.\ D {\bf 72}, 015004 (2005)
  [hep-ph/0504036].

\bibitem{Falkowski:2005ck} 
  A.~Falkowski, O.~Lebedev and Y.~Mambrini,
  JHEP {\bf 0511}, 034 (2005)
  [hep-ph/0507110].

\bibitem{Kachru:2003aw} 
  S.~Kachru, R.~Kallosh, A.~D.~Linde and S.~P.~Trivedi,
  Phys.\ Rev.\ D {\bf 68}, 046005 (2003)
  [hep-th/0301240].
  
\bibitem{Baer:2006tb} 
  H.~Baer, E.~K.~Park, X.~Tata and T.~T.~Wang,
  Phys.\ Lett.\ B {\bf 641}, 447 (2006)
  doi:10.1016/j.physletb.2006.08.075
  [hep-ph/0607085].
  
\bibitem{Baer:2007eh} 
  H.~Baer, E.~K.~Park, X.~Tata and T.~T.~Wang,
  JHEP {\bf 0706}, 033 (2007)
  doi:10.1088/1126-6708/2007/06/033
  [hep-ph/0703024].
  
\bibitem{Kaufman:2013oaa} 
  B.~Kaufman and B.~D.~Nelson,
  Phys.\ Rev.\ D {\bf 89}, no. 8, 085029 (2014)
  doi:10.1103/PhysRevD.89.085029
  [arXiv:1312.6621 [hep-ph]].
  
  
\bibitem{Matalliotakis:1994ft} 
  D.~Matalliotakis and H.~P.~Nilles,
  Nucl.\ Phys.\ B {\bf 435}, 115 (1995)
  doi:10.1016/0550-3213(94)00487-Y
  [hep-ph/9407251].
  P.~Nath and R.~L.~Arnowitt,
  Phys.\ Rev.\ D {\bf 56}, 2820 (1997)
  doi:10.1103/PhysRevD.56.2820
  [hep-ph/9701301].
  J.~R.~Ellis, K.~A.~Olive and Y.~Santoso,
  Phys.\ Lett.\ B {\bf 539}, 107 (2002)
  doi:10.1016/S0370-2693(02)02071-3
  [hep-ph/0204192].
  J.~R.~Ellis, T.~Falk, K.~A.~Olive and Y.~Santoso,
  Nucl.\ Phys.\ B {\bf 652}, 259 (2003)
  doi:10.1016/S0550-3213(02)01144-6
  [hep-ph/0210205].
  H.~Baer, A.~Mustafayev, S.~Profumo, A.~Belyaev and X.~Tata,
  JHEP {\bf 0507}, 065 (2005)
  doi:10.1088/1126-6708/2005/07/065
  [hep-ph/0504001].
  
  \bibitem{Everett:2008qy} 
  L.~L.~Everett, I.~W.~Kim, P.~Ouyang and K.~M.~Zurek,
  Phys.\ Rev.\ Lett.\  {\bf 101}, 101803 (2008)
  [arXiv:0804.0592 [hep-ph]].
 
\bibitem{Everett:2008ey} 
  L.~L.~Everett, I.~W.~Kim, P.~Ouyang and K.~M.~Zurek,
  JHEP {\bf 0808}, 102 (2008)
  [arXiv:0806.2330 [hep-ph]].
 
 
\bibitem{Altunkaynak:2010tn} 
  B.~Altunkaynak, B.~D.~Nelson, L.~L.~Everett, Y.~Rao and I.~W.~Kim,
  Eur.\ Phys.\ J.\ Plus {\bf 127}, 2 (2012)
  [arXiv:1011.1439 [hep-ph]].
 
\bibitem{Everett:2015dqa} 
  L.~L.~Everett, T.~Garon, B.~L.~Kaufman and B.~D.~Nelson,
  [arXiv:1510.05692 [hep-ph]].
  
  
\bibitem{Abe:2014kla} 
  H.~Abe and J.~Kawamura,
  JHEP {\bf 1407}, 077 (2014)
  [arXiv:1405.0779 [hep-ph]].

\bibitem{Allanach:2001kg} 
  B.~C.~Allanach,
  Comput.\ Phys.\ Commun.\  {\bf 143}, 305 (2002)
  [hep-ph/0104145].

\bibitem{Ade:2013zuv} 
  P.~A.~R.~Ade {\it et al.}  [Planck Collaboration],
  [arXiv:1303.5076 [astro-ph.CO]].

\bibitem{Belanger:2001fz} 
  G.~Belanger, F.~Boudjema, A.~Pukhov and A.~Semenov,
  Comput.\ Phys.\ Commun.\  {\bf 149}, 103 (2002)
  [hep-ph/0112278].


\bibitem{ATLAS:2013mma} 
  [ATLAS Collaboration],
  ATLAS-CONF-2013-014.
\bibitem{Chatrchyan:2013lba} 
  S.~Chatrchyan {\it et al.}  [CMS Collaboration],
  JHEP {\bf 06}, 081 (2013)
  [arXiv:1303.4571 [hep-ex]].
\bibitem{Aad:2015zhl} 
  G.~Aad {\it et al.} [ATLAS and CMS Collaborations],
  Phys.\ Rev.\ Lett.\  {\bf 114}, 191803 (2015)
  [arXiv:1503.07589 [hep-ex]].


\bibitem{Chatrchyan:2013bka} 
  S.~Chatrchyan {\it et al.} [CMS Collaboration],
  Phys.\ Rev.\ Lett.\  {\bf 111}, 101804 (2013)
  [arXiv:1307.5025 [hep-ex]].
\bibitem{Aaij:2012nna} 
  R.~Aaij {\it et al.} [LHCb Collaboration],
  Phys.\ Rev.\ Lett.\  {\bf 110}, no. 2, 021801 (2013)
  [arXiv:1211.2674 [hep-ex]].
\bibitem{Aaij:2013aka} 
  R.~Aaij {\it et al.} [LHCb Collaboration],
  Phys.\ Rev.\ Lett.\  {\bf 111}, 101805 (2013)
  [arXiv:1307.5024 [hep-ex]].
  
\bibitem{Asner:2010qj} 
  D.~Asner {\it et al.} [Heavy Flavor Averaging Group Collaboration],
  [arXiv:1010.1589 [hep-ex]].

\bibitem{Aad:2015baa} 
  G.~Aad {\it et al.} [ATLAS Collaboration],
  JHEP {\bf 1510}, 134 (2015)
  [arXiv:1508.06608 [hep-ex]].
  
\bibitem{Cohen:2013xda} 
  T.~Cohen, T.~Golling, M.~Hance, A.~Henrichs, K.~Howe, J.~Loyal, S.~Padhi and J.~G.~Wacker,
  JHEP {\bf 1404}, 117 (2014)
  [arXiv:1311.6480 [hep-ph]].

\bibitem{Baer:2011ab} 
  H.~Baer, V.~Barger and A.~Mustafayev,
  Phys.\ Rev.\ D {\bf 85}, 075010 (2012)
  [arXiv:1112.3017 [hep-ph]].

\bibitem{Brummer:2012ns} 
  F.~Brummer, S.~Kraml and S.~Kulkarni,
  JHEP {\bf 1208}, 089 (2012)
  [arXiv:1204.5977 [hep-ph]].


\end{thebibliography}
\end{document}